# Relaxation Control of Packet Arrival Rate in the Neighborhood of the Destination in Concentric Sensor Networks

[1]T.R.Gopalakrishnan Nair (SM-IEEE), [2]R. Selvarani, [3]Vaidehi M.
[1]Director – Research & Industry Incubation Centre
[2]Associate Professor – Research –CSE
Research Associate
[3]RIIC, Dayananda Sagar Institutions, Bangalore-78
vaidehidm@yahoo.co.in

## Abstract

**One of the challenges in the wireless sensor applications which are gaining much attention is the real-time transmission of continuous data packets across the network. Though advances in communication in sensor networks are providing guaranteed quality data packet delivery they still have some drawbacks. One such drawback is transmission of incessant data packets over high speed networks. Here in this paper we have designed a concentric sensor network having buffer just not at the sink but also in selected intermediate nodes to minimize the packet loss caused due to congestion. This approach results in haggle congestion and less packet loss in the designed network**.

## I. Introduction

In a sensor network the nodes are spatially distributed. These nodes or sensors work in a coordinated mode to monitor physical or environmental changes like variation in temperature – prediction of a forest fire, vibration – occurrence of an earthquake and prediction of landslides etc. These networks are now employed in habitable monitoring, health care applications and traffic control.

The nodes are deployed according to the network topology, geographical conditions and the nature of application. These nodes in the network are equipped with wireless communication device or transceiver microcontrollers, and a battery as an energy source. This network is usually comprised of wireless ad-hoc system meaning that each sensor supports a multihop routing algorithm. The information is transmitted in bit streams and later is packetized. This causes some packetization delay at the nodes in the network that collects information and transmits to the sink node for further processing.

The performance of the network is estimated based on the rate of delivery of the packets to the desired node in the network. There are various parameters which affect the quality of a sensor network among which jitter is one of the factor which leads to loss of synchronization of packets down stream from an active node. As the packets have to be transmitted through a sequence of nodes jitter accumulates and intensifies in the network leading to delay in delivery of the packets and which might cause congestion. Congestion in a network will lead to dropping of packets, at times some of the high priority packets will also be lost. Congestion also increases the consumption of energy as links become saturated.
In this paper we present a network topology with a buffer integrated with selected nodes to minimize jitter in the sensor network. Section II presents the literature survey on Quality of service and the performance metrics applied to estimate the network efficiency. The problem definition, the design of the network and related analysis are presented in section III. Section IV is about conclusion and future work.

## II. Quality of Service Metrics

The set of service requirements that need to be provided by the network at the time of transferring packets from the source to the destination reflects on the performance of the network. The network is expected to assure a set of measurable service attributes to the user in terms of bandwidth, probability of packet loss, energy and delay variation, and end to end delay [1, 6]. The QoS metrics are classified as additive metrics, concave metrics, and multiplicative metrics.Bandwidth and energy are considered as concave metrics while cost, delay and jitter as to additive metrics. The end-to-end delay is an additive constraint because it is the accumulation of all delays of the links along the path and link the break probability is a multiplicative metric [2].

### *A. Performance metrics*

Several performance metrics are evaluated under various mobility speeds on the following calculation [4].

- Throughput

  The amount of data that is received through the network per unit time, i.e, data bytes delivered to their destination per second.

$$Throughput = \frac{Total\ bytes\ received}{Total\ Time}$$

- End-to-end delay

  This indicates the time taken for a packet to travel from CBR (constant bit rate source) to the application layer of the destination.

  It represents the average data delay of an application or a user experiences while transmitting a data packet.

- Jitter

  It is the time variation of a characteristic of the latency packets at the destination.

- Packet delivery fraction

$$packet\ delivery\ fraction = \frac{total\ received\ packets}{sent\ packets} * 100$$

## III. Problem Definition and Design Architecture

*Problem definition*

Jitter is one of the parameters which affect the performance of the sensor network leading to a delay in packet delivery and also the loss of the packets.The Figure1 shows the solution adopted to minimize the delay in the network. Here we have incorporated a preemptive FIFO stack in the transport layer. In this buffer the packets are first accumulated and synchronized [7, 8] before transmitting to the next intermediate node or to the sink. If there is a packet in the buffer, it will be preempted by the newly arriving high priority packet (router).

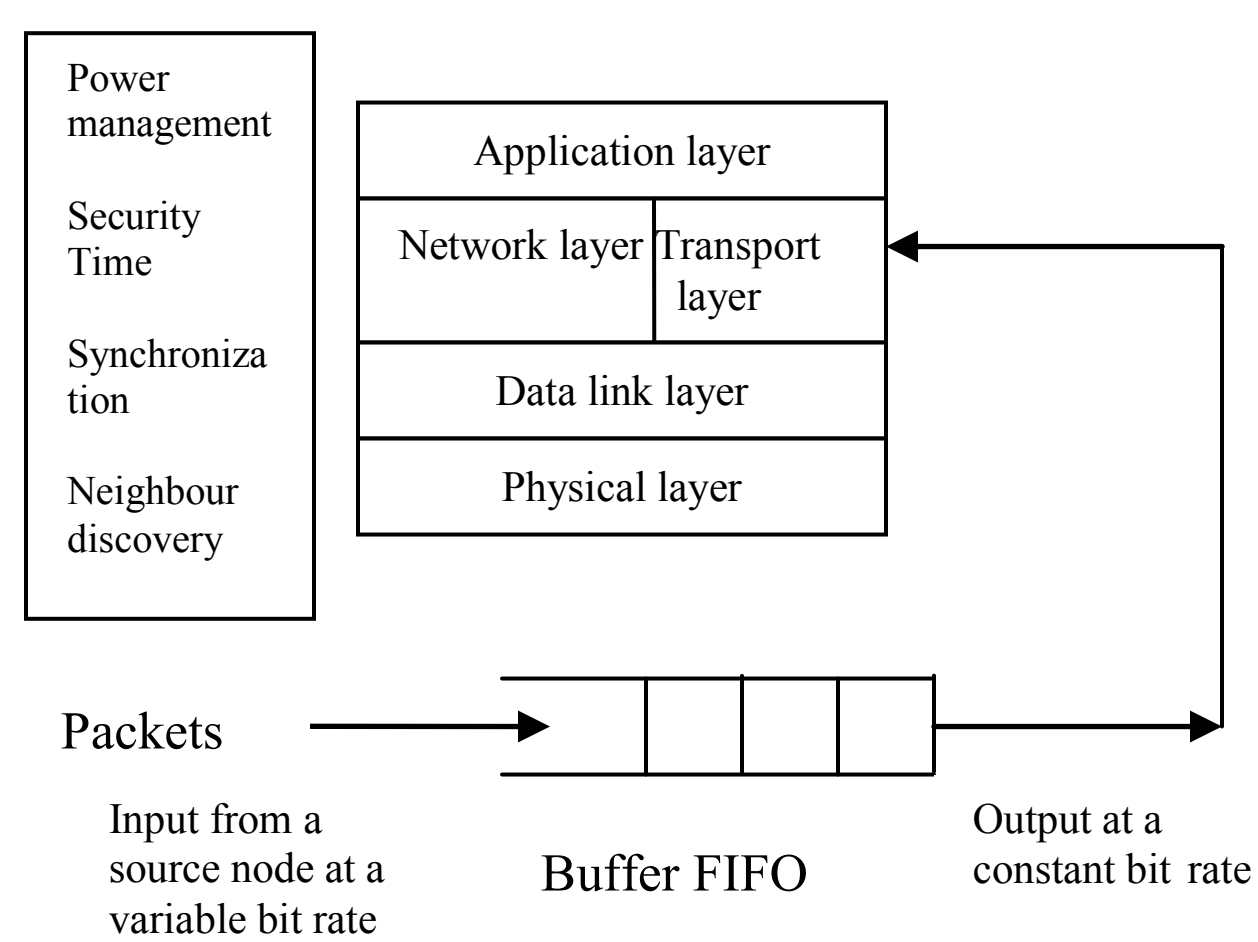

**Figure 1. Design architecture**

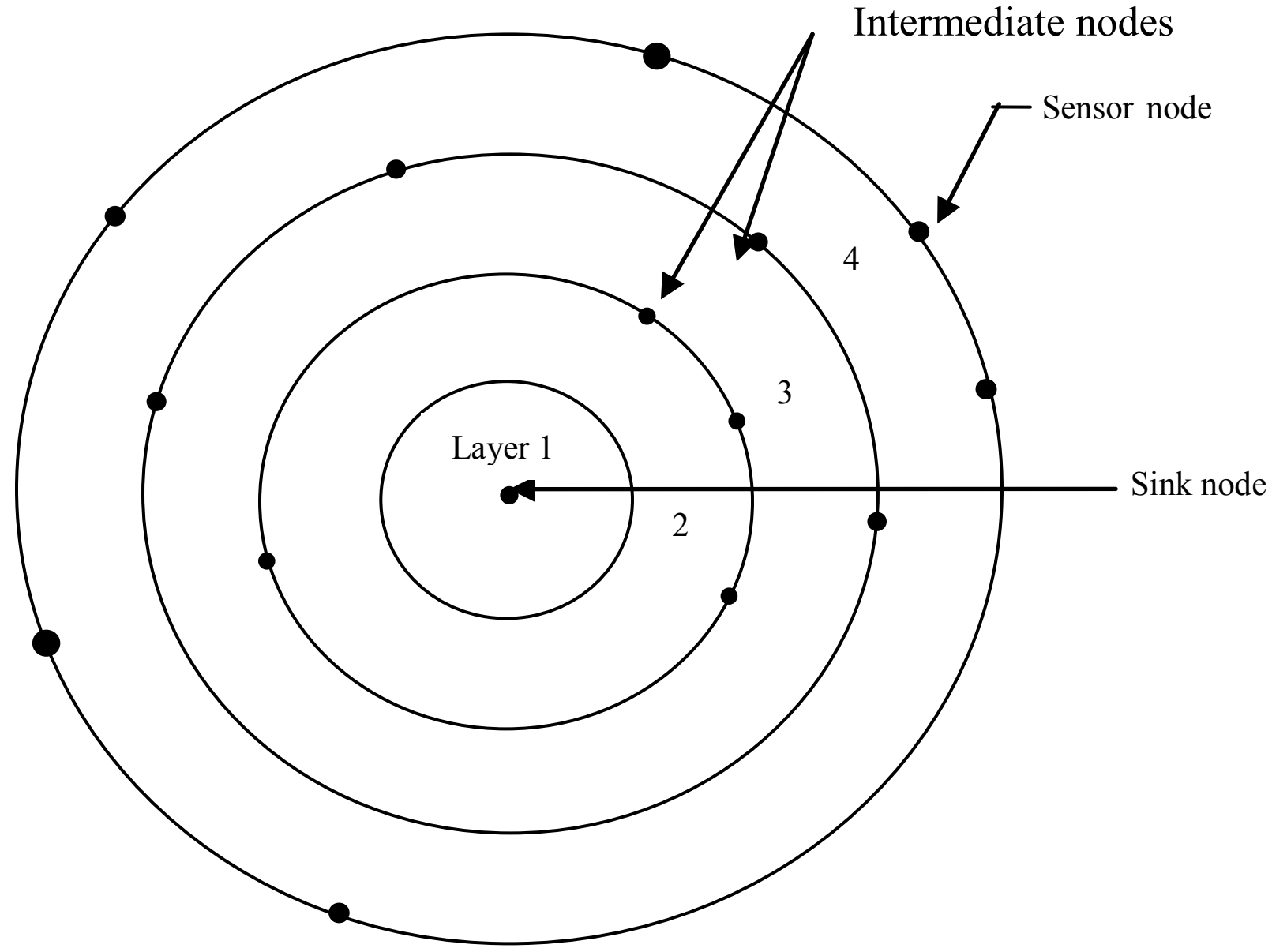

**Figure 2. Network Topology**

*Buffer size of selected node* $\propto$ *(4-n)×packet size.*

for n<4,

for n> 4, Buffer size = 1.

n= layer in the network…..Eq. 1

The Figure 2 shows the topology, to minimize the cost and the delay, selected nodes [9] are integrated with the buffer.

$T_l$ = life time of a packet

$P_i$ = prioritized packets

If the packet is ahead of schedule, it is held just long enough to get it back on schedule [3]. Here, the packet is behind the schedule due to the packetization delay. The intermediate node having a buffer tries to send the packet at the earliest. Steps to be followed by the overcrowded node with a buffer integrated in it

- If intermediate node is overcrowded with undelivered packets then
- For a packet $P_n$ check the life time and packet priority.
- if $T_l$ <= 0 && Pi is high then preempt the packets in buffer then transmit the packet $P_n$

Assumptions

1. The bits arrive at variable bit rate. Prior to transmission towards the destination node, they are packetized. The buffer output will be of CBR causing packetization delay shown in figure 3.

2. Buffer size is fixed.

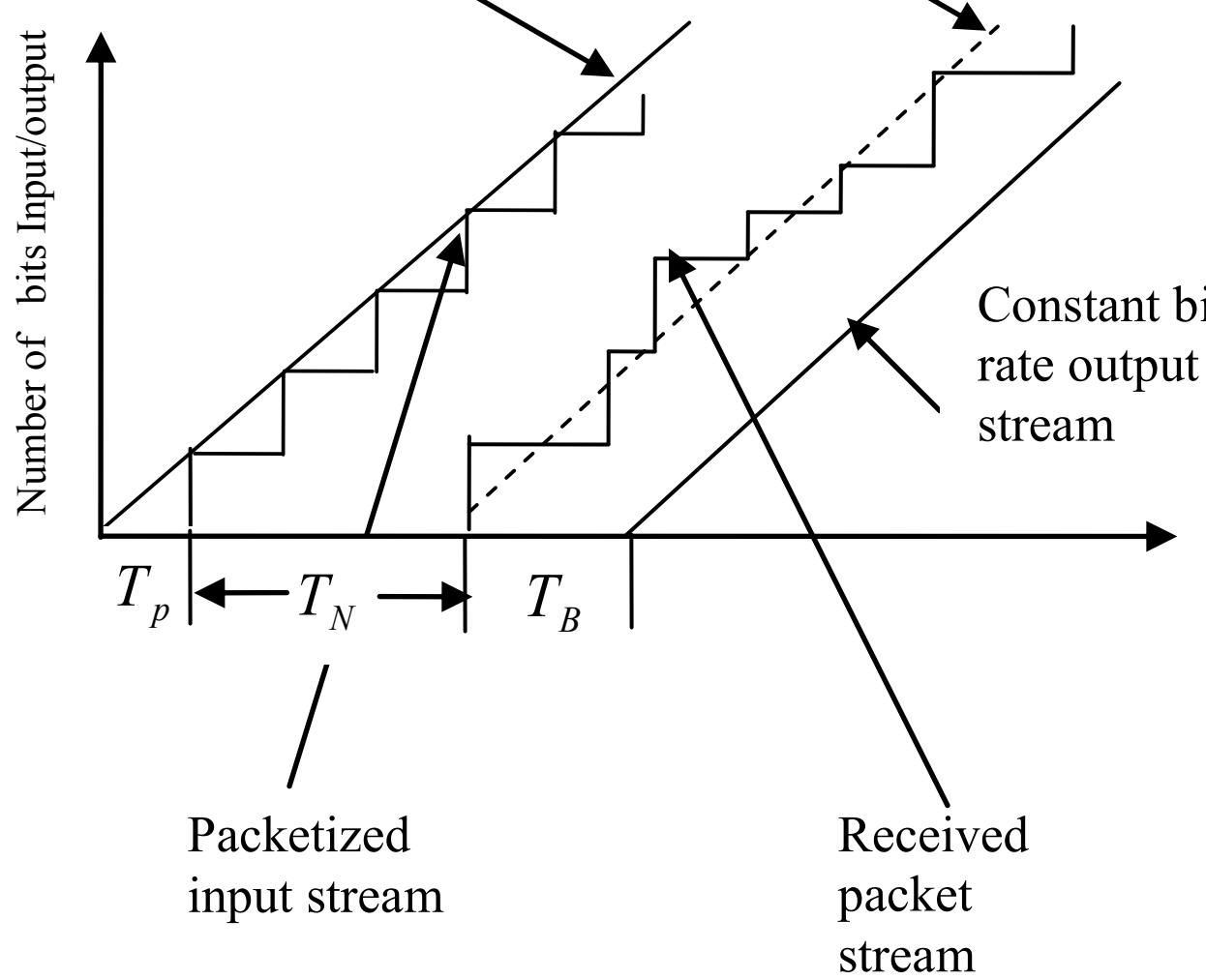

**Figure 3. Transmission of a constant bit rate stream** [2]

$T_P$= packetization delay
$T_N$ = mean network packet transfer delay
$T_B$ = buffering delay at destination (to overcome worst-case jitter)
$T_t$ = total input-output delay
$T_t = Tp + T_N + T_B$
Jitter = variation in store-and-forward delay about the mean.

## Performance Evaluation

**Table 1. Simulation Parameters**

| | |
|---|---|
| Simulation area | $2500000m^2$ |
| Number of nodes | 100 |
| Node communication range | 50m |
| Traffic model | CBR |
| Packet size | 512bytes |
| Node Energy | 100J |

The transmitted packet size and the buffer size are the prime parameters considered to evaluate jitter. From Table 2 it is observed that when in the absence of a buffer at any node in the communication channel, the jitter associated with the packets are increased with the increase in the number of packets transmitted.

This may lead to loss in the data to be reached to the destination. This scenario is graphically depicted in Figure 4.

**Table 2.Jitter without buffer at intermediate nodes**

| Delay in ms | No. of packets | Jitter in ms |
|---|---|---|
| 0.2 | 10 | 1 |
| 0.4 | 20 | 2 |
| 0.6 | 30 | 3 |
| 0.8 | 40 | 4 |
| 0.9 | 50 | 5 |
| 1 | 60 | 6 |

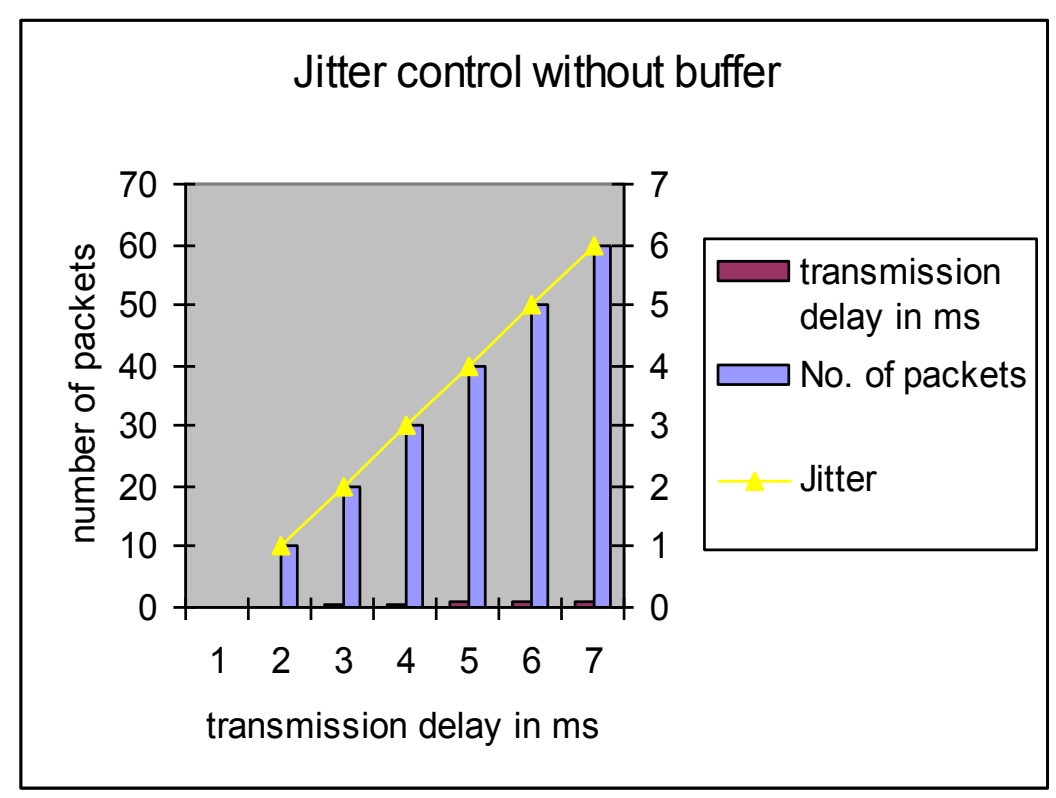

**Figure 4. Case without jitter control**

**Table 3. Jitter buffer at intermediate nodes**

| Delay due to packetization ms | No. of packets | Jitter in ms |
|---|---|---|
| 3.5 | 10 | 5 |
| 4 | 20 | 5 |
| 4.5 | 30 | 5 |
| 5 | 40 | 5 |
| 5.5 | 50 | 5 |

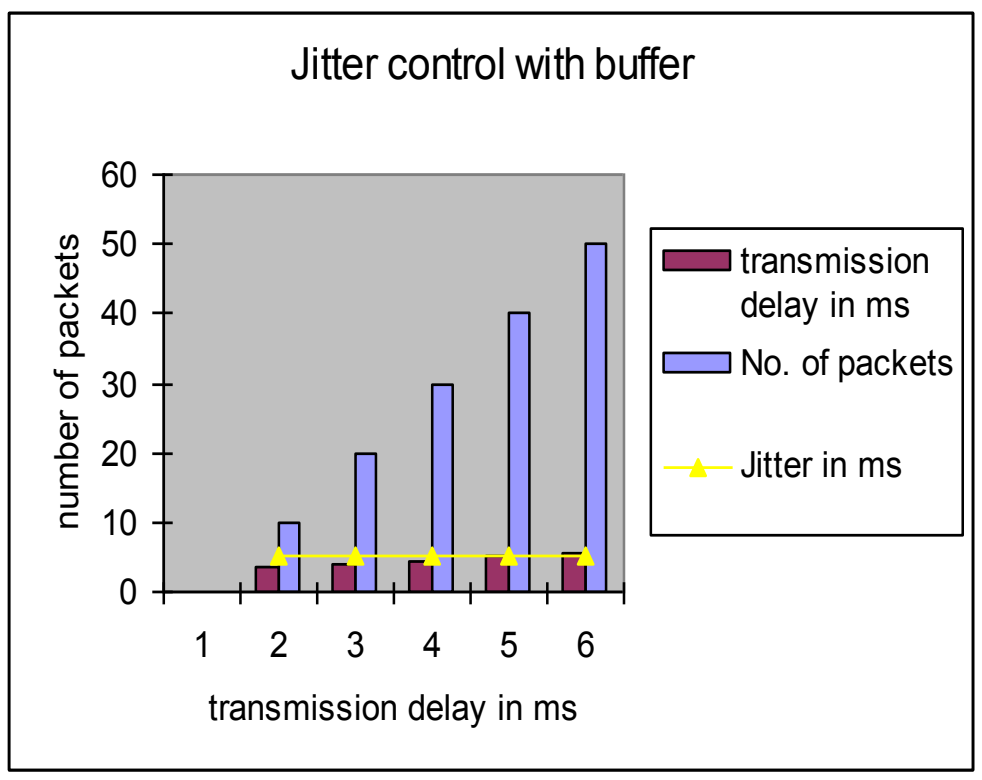

**Figure 5. Case with jitter control**

On observing the Table 3 it is evident that on including the buffer size as estimated using Eq. 1 at selected node in the communication channel of the sensor network the jitter associated with the packets arrived can be controlled to have a constant value of jitter of 05ms. The packet stream with 50 packets is having a jitter of about 05ms Figure 5 depicts the graphical representation of the above situation.

Though the over all delay is increased due to packetization and also due to the store and forward process at the buffer, there is no packet loss and also congestion is under control.

## IV. Conclusion and future work

Our solution does not require an active queue management, maintenance of multiple queues or the use of a specified MAC protocol. It is for real time application for which we have used a scheduling technique where in the high priority packets are transmitted first. By introducing the buffer we have fixed the jitter to a particular level, though the delay is increased due to the packetization process the packets are not lost. In our future work we would like to achieve the technique of identifying the intermediate nodes to introduce the buffer to enable synchronization of the packets.

## References

[1] N. Sengottaiyan, R M.Somasundaram, S. Arumugam, "*A Modified Routing Algorithm for Reducing Congestion in Wireless Sensor Networks*", European Journal of Scientific Research, Euro Journals Publishing, Inc. 2009, ISSN 1450-216X Vol.35 No.4 (2009), pp.529-536.

[2] Fred Halsall, Multimedia communications Applications, Networks, Protocol and Standards, Pearson Education (Singapore) Pte. Limited, 2002

[3] R. Asokan, A.M. Natarajan and C.Venkatesh "*Ant Based Dynamic Source Routing Protocol to Support Multiple Quality of Service (QoS) Metrics in Mobile Ad Hoc Networks*" Journal of Computer Science and Security, volume (2) issue (3).

[4] Jane W.S. Liu, Real-Time Systems, Pearson Education, Inc. and Dorling Kindersley Publishing Inc.

[5] Boaz Patt, Shamir Mansour, "Jitter control in QoS networks" IEEE/ACM Transactions on Networking (TON) Volume 9, issue 4 (August 2001) table of contents Pages: 492 – 502 Year of Publication: 2001 ISSN:1063-6692

[6] Dr.P.Bala Krishna Prasad Mr.G.Murali G.Gurukesava Das, K.Siva Krishna Rao, "*Congestion Controlling for Streaming Media Through Buffer Management and Jitter Control*", IJCSNS International Journal of

Computer Science and Network Security, VOL.9 No.2, February 2009.

[7] Dong Zhou, Ten H. Lai , *"An Accurate and Scalable Clock Synchronization Protocol for IEEE 802.11-Based Multihop Ad Hoc Networks",* IEEE Transactions on Parallel and Distributed Systems Volume 18 , Issue 12 (December 2007), Pages 1797-1808, Year of Publication: 2007 , ISSN:1045-9219.

[8] Dong Zhou, Ten H. Lai, *"A Scalable and Adaptive Clock Synchronization Protocol for IEEE 802.11-Based Multihop Ad Hoc Networks"*, Publication Date: 7-7 Nov. 2005, on page(s): 8 pp.-558, ISBN: 0-7803-9465-8.

[9] Hu Zhang, Huiyan Zhang, *"Node Selection Algorithm Optimized for Wireless Sensor Network"*, ISBN: 0-7695-3090-7, First International Workshop on Knowledge Discovery and Data Mining 2008.